\documentstyle[prl,aps,graphicx]{revtex}

\newcommand{\be}{\begin{equation}} \newcommand{\ee}{\end{equation}}
\newcommand{\ba}{\begin{array}} \newcommand{\ea}{\end{array}}
\begin{document} \draft 
\title{One relation for self-gravitating bodies}
\author {Zakir F.Seidov and P.I. Skvirsky}  
\address{Department of Physics, Ben-Gurion University of the Negev \\
P.O. Box 653, Beer-Sheva 84105, Israel\\
email:seidov@bgumail.bgu.ac.il} \maketitle 
\begin{abstract} The relation between the gravitational potential energy,
the central potential, and the mass is considered for various
self-gravitating bodies.\end{abstract}
\pacs{PACS numbers: 01.55.+b, 45.20.Dd, 96.35.Fs}
\section{Introduction}
For homogeneous triaxial ellipsoid with semiaxes $A,\,B,\,C\,$ and density $\rho$ the
next relations are valid (Landau and Lifshits 1975).
The gravitational potential  at the inner point ($X,\,Y,\,Z\,$), 
$-A\leq X \leq A$, $-B\leq Y \leq B$, $-C\leq Z \leq C$, is 
\be \label{Uel} \ba{l}U_{ell}(X,Y,Z)=\pi\,\rho\,G\,A\,B\,C\,\int_0^\infty \,
\biggl(1 -{X^2 \over A^2+s}-{Y^2 \over B^2+s}-{Z^2\over C^2+s} \biggr)
{\,d\,s \over Q_s}; \\Q_s=\sqrt{(A^2+s)(B^2+s)(C^2+s)}. \ea \ee 
The potential energy of the homogeneous triaxial ellipsoid  is: 
\be \label{Wel} W_{ell}={3\over 10}G\,M^2\int_0^\infty{d\,s\over Q_s}.\ee
Here $M_{ell}=4/3\,\pi\,\rho\,A\,B\,C $ is the ellipsoid's mass and $G$ stands for
Newtonian constant of gravitation. Note that gravitational energy of self-gravitating body
is of negative sign but we loosely write all $W$s with positive sign.
In general case $A\neq B \neq C$ the integrals in (\ref{Uel}) and (\ref{Wel})
are expressed only in terms of the incomplete elliptic integrals and this
precludes any detailed analysis. \
However, if we consider only potential at the center
of ellipsoid, $U_{ell}(0,0,0)$, then we get the remarkable relation:
\be \label{WUMel} WUM={W_{ell} \over U_{ell}(0,0,0)\,M_{ell}}= {2\over 5}, \ee
valid for any values of semi-axes. 
We shortly refer to this relation (\ref{WUMel}) as $WUM$-ratio.\\
Recently Seidov and Skvirsky (2000a)  presented the gravitational potential and the
potential energy for  the homogeneous rectangular parallelepiped (hereafter RP) which
allows to analyse  this $WUM$-ratio for the new class of the homogeneous self-gravitating
bodies.\\
	In this paper, we show in sections II and III that the value of
$WUM$ for RPs has minimal value $.395437$ (see Eq. (\ref{WUMC})) for the cube  (all three
dimensions of RP equal to each other), tends to 1/2 for one dimension of RP far larger 
larger than two others (long thin "stick" with square cross-section), and tends to 
{${1\over 2}-{\sqrt{2}-1\over 6\,\ln(\sqrt{2}+1)}=.421673$} as one dimension is far
less than two others (thin square "plate"), see Fig. 1.\\
Also, in section \ref{KANT} we discuss the $WUM$-ratio for homogeneous gravitating bodies
studied recently by Kondrat'ev and Antonov (1993). We show that values of $WUM$-ratio for
the homogeneous summetrical lenses are in the interval from $128/105\,\pi$, for 
infinitesimally thin symmetrical lens,  to $17/20$, for two homogeneous 
equal spheres just touching each other.\\ In section \ref{poly} we analyse $WUM$-ratio for
spherical polytropic stars, and show that $WUM$-ratio varies from $2/5$ to $3/32 \,\pi$
for  polytropic index $n$ varying from $0$ to $5$.\\  
In the section \ref{2ph} we discuss the interesting class of two-phase  spheres and show
 that unlike the polytropes, in this case the $WUM$-ratio's interval is larger: it is
possible to get very small values of  $WUM$ if the ratio of two densities  $q=\rho_2/\rho_1$ 
is large enough and if the relative value of core's radius is not too  small. \\  
At last, in the sections \ref{step} and \ref{alpha}  we consider another two simple
classes of {\it non-homogeneous} bodies both allowing analytical treatment. 
\section{Potential at the center of RP} \label{SURP}
Using results by Seidov and Skvirsky (2000a)  we write down the gravitational potential at
the center of the homogeneous RP  with density $\rho$ and with edge lengths
$2a,\,2b,\,2c$: \be \label{URP} \ba{l} U_{RP}(0,0,0)=4\,G\,\rho\,
[ a\, b \ln {d+c\over d-c}+ b\, c \ln {d+a\over d-a}+ c d \ln {d+b\over d-b}-\\ \\
a^2 \arctan {b\, c \over a\, d}- b^2 \arctan {a\, c \over b\, d}- 
c^2 \arctan {a\, b \over c\, d}].\ea \ee
Here $d=(a^2+b^2+c^2)^{1/2}$ is the main diagonal of RP. \\
Three particular cases are of the larger interest:\\
a){\it cube} corresponding to case {$c=b=a$}, 
\be \label{p000} U_{cube}(0,0,0)= G\,\rho\,a^2
\biggl[24 \ln{1+\sqrt{3}\over\sqrt{2}}-2\pi\biggr] = 9.52017\,G\,\rho\,a^2;\ee
b){\it long thin stick with square cross-section} corresponding to case $a>>b=c$:
\be \label{Ust} U_{stick}(0,0,0)=
G\,\rho\,b^2(-8\,\ln {b\over a}+12-2\,\pi+4\,\ln 2);\ee
c){\it thin square plate} corresponding to the case $a<<b=c$:
\be \label{Upl} \,U_{plate}(0,0,0)=G\,\rho\,[16\,a\,b\,\ln (\sqrt{2}+1)-2\,\pi \,a^2].\ee
\section{Potential energy of RP} According to Seidov and Skvirsky (2000a) the
gravitational potential energy of the homogeneous rectangular parallelepiped is equal to:
\be \label{WRP} \ba {l} WRP=G\,\rho^2 [f(a,b,c)+f(b,c,a)+f(c,a,b)];
f(a,b,c)=c_5 a^5+c_4 a^4+c_3 a^3+c_2 a^2; \\ \\
 c_5={32\over 15};\,\,c_4={32\over 15}(d-d1-d3)-
{16\,b\over 3}\ln{(d1-b)(d+b)\over a\,d3}-
{16\,c\over 3}\ln{(d3-c)(d+c)\over a\,d1};\\ \\
c_3=-{64\, b\, c\over 3}\arctan {b\, c\over a\,d};\quad 
d=\sqrt{a^2+b^2+c2};\,d1=\sqrt{a^2+b^2};\,d3=\sqrt{a^2+c^2};\\ \\
c_2={32\,b^2 \over 5} ( d1 -d)  + {32\,c^2  \over 5} ( d3 -d ) - 
 16\,b\,{c^2}\,\ln {d-b\over d+b}-16\,{b^2}\,c\,\ln {d-c\over d+c}.\ea\ee
\subsection{Potential energy and $WUM$-ratio of cube}
From (\ref{WRP}), taking $c=b=a$, we get the potential energy of homogeneous
cube with edge length $2 a$:
\be \label{WC} \ba{l} W_{cube}=32 G\rho^2 a^5 \{{2\sqrt{3}-\sqrt{2}-1\over
5}+{\pi\over3}+ \ln [(\sqrt{2}-1)(2-\sqrt{3})]\}= 30.117\, G \rho^2 a^5. \ea \ee 
From (\ref{p000}) and (\ref{WC}) we get $WUM$-ratio for homogeneous cube:
\be \label{WUMC} WUM_{cube}=2 {{2\sqrt{3}-\sqrt{2}-1\over
5}+{\pi\over 3}+\ln [(\sqrt{2}-1)(2-\sqrt{3})]\over
24 \ln{1+\sqrt{3}\over\sqrt{2}}-2\pi}=.395437.\ee
\subsection{Potential energy and $WUM$-ratio of thin long stick}
We take $a>>b=c$ that corresponds to the  case of the thin long stick with the square
cross-section. Leading term in expansion of WRP (\ref{WRP}) gives the 
potential energy of the thin long stick:
\be \label{Wst} W_{stick}={32\over 3}G \rho^2 a\,b^4\ln{a\over b}. \ee
From this and (\ref{Ust}) we get for stick:
\be WUM={W_{stick}\over 8\,a\,b^2 U_{stick}(0,0,0)}={1\over 2} .\ee
\subsection{Potential energy and $WUM$-ratio of thin square plate}
Taking one of RP's dimension infinitesimally small, $a\rightarrow 0$, we get, 
from Eq.(\ref{WRP}), the potential energy of the thin rectangular plate. If we
additionally take $b=c$, then we get the potential energy of the thin square  plate
($a<<b$): \be \label{Wpl} W_{pl}=64\,G\,\rho^2b^3\,a^2\,\left(\ln (\sqrt{2}+1)-
{\sqrt{2} -1 \over 3}  \right) =47.5714\, G\,\rho^2b^3 a^2.\ee
From (\ref{Upl}) and (\ref{Wpl}) we have another limit for value of $WUM$:
\be \label{WUMpl} {WS\over 8\,a\, b^2 U_{pl}(0,0,0)}=
{1\over 2}-{\sqrt{2}-1\over 6\,\ln(\sqrt{2}+1)}=.421673.\ee
\begin{figure}  \label{RPfig} \includegraphics[scale=.6]{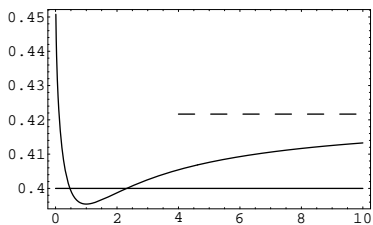}
\caption{WUM for RP with square cross section $2\,b\,$x$\,2\,b$ and length $2\,a$.
Abscissas are values of $b/a$ and ordinates are values of relation $WUM=W/U(0)\,M$ that is
ratio of gravitational potential energy of RP to product of gravitational  potential at
the center of RP by mass of homogeneous RP. $WUM$ has minimum for cube ($b/a=1$), tends to
1/2 at $b/a\rightarrow 0$ (thin long stick) and to
${1\over 2}-{\sqrt{2}-1\over 6\,\ln(\sqrt{2}+1)}=.421673$ (dash line ) at 
$b/a\rightarrow \infty$ (thin square plate).
For homogeneous ellipsoid, $WUM=2/5$, solid line.} \end{figure} 
\subsection{Relation between potential energy, gravitational potential and mass of RP}
General behavior of relation between the potential energy, the gravitational potential
at the center, and mass of the homogeneous RP with two equal edge-lengths is shown in the
Fig. 1, which is a result of numerical calculation by  the formulas
(\ref{URP}) and (\ref{WRP}).  For homogeneous ellipsoid $WUM=2/5$, see (\ref{WUMel}),
solid line in Fig. 1.  
\section{Homogeneous symmetric lenses} \label{KANT}
Recently Kondrat'ev and Antonov (1993) (hereafter KA) have obtained the
 analytical formulas for the gravitational potential and the
gravitational energy of some  axial-symmetric figures,  namely
homogeneous  lenses with spherical surfaces of different radii. In
forthcoming paper (Seidov and Skvirsky 2000b) we present some new
solutions for homogeneous bodies of revolution. \\
Here we present  the review of $WUM$ for most suitable kind of those bodies, discussed by
KA, namely the homogeneous symmetrical lenses. A segment of sphere, or a planoconvex lens,
is obtained by cutting a sphere with a plane. If $R$ is a radius of a sphere,  $h$ is
height of segment, and $2\,a$ is a radius of segment's base, then $a=\sqrt{ 2\,h\,R-h^2}$.
A symmetric homogeneous lens (SL) is obtained by placing together the bases of two
identical segments of sphere.\\ 
According to KA we have for the gravitational potential at the center of such SL: 
\be \label{USL} U_{SL}(0)={4\,\pi \,G\,\rho\over 3}
\left(  h\,R  + R^2 -{h^2\over 2}  + {R^3 - a^3 \over h - R} \right) .\ee
The gravitational energy of SL is:
 \be \label{WSL} \ba{l} W_{SL}={4\,\pi \,G\,{\rho}^2\over 9}
     \biggl[10\,R^4\,a - {16\over 3}\,R^2\,a^3- {8\over 3}\,a^5+
       \pi\,{h^4}\,( {2\over 5}\,h - 2\,R - {R^2\over R - h } )+ 
       2\,{R^4}\,\biggl( {R^2\over  R - h} - 
          6\,(R -h)\biggl) \, \arctan {a\over R-h })\biggr]. \ea \ee
And the total mass of the homogeneous SL is:
\be \label{MSL} M_{SL}={2\,\pi \,\rho\,h^2\over 3}\,(  3\,R -h). \ee
\begin{figure}  \label{SLfig} \includegraphics[scale=.6]{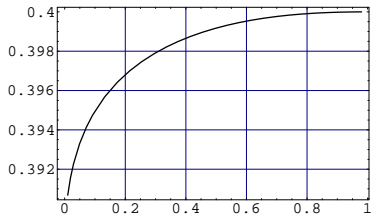}
\caption{WUM for SL.  Abscissas are values of $h/R$, ratio of height of half-lens to
radius of sphere and ordinates are values of relation $WUM=W/U(0)\,M$.
At $h/R\rightarrow 0$ (thin spherical lense),  
$WUM\rightarrow {128\over 105\,\pi}=.388035$. At $h/R\rightarrow 1$,
 (full sphere) $WUM\rightarrow 2/5$.} \end{figure} 
We may consider $WUM$-ratio for such bodies as $W_{SL}/U_{SL}\,M_{SL}$.
Defined so, $WUM$ for the homogeneous symmetric lenses has general
dependence on parameter $h/R$, according the formulas (\ref{USL}), 
(\ref{WSL}) and (\ref{MSL}), as shown in the Fig. 2. Note that in the
 limit $h/R \rightarrow 0$ we have infinitesimally thin symmetrical lens
with radius of curvature tending to infinity, and we have: 
\be Limit[WUM_{SL}]_{h/R\,->0}={128\over 105\,\pi}=.388035.\ee
Interestingly, this infinitesimally thin round lens does not coincide at all 
with the case of the infinitesimally thin  quadratic  plane  which has the 
much more larger value of  $WUM=.421673$ (see Eq.(\ref{WUMpl})).
In another limit $h/R \rightarrow 1$ we have a full homogeneous sphere,  and evidently:
\be Limit[WUM_{SL}]_{h/R->1}=2/5.\ee
There is still another $analytical$ case at $h\rightarrow 2\,R\,$ when we 
have two homogeneous spheres touching each other so that distance between
centers of spheres is $d=2\,R;$ then "potential at the center of SL" is
$U_{SL}(0)=2\,{G\,M\over R}$; potential energy is sum of two terms: proper
potential energy of each spheres $2\,{3\over 5}\,{G\,M^2\over R}$, and
$interaction$ energy, ${G\,M^2\over d}$ or  ${G\,M^2\over 2\,R }$; we have:
$$WUM_{SL}|_{h=2\,R}=({1\over 2}+{6\over 5})/2={17\over 20}.$$ 
\section{Heterogeneous spherical bodies} \label{hetero}
Now we consider the problem of $WUM$-ratio from another point of view.
Abovementioned {\it homogeneous} gravitating bodies (ellipsoids, right parallelepipeds,
 and symmetrical spherical lenses) differ from each other only {\it by their forms} 
and so $WUM$-ratio may be referred to as {\it form-factor}. As it is evident, $WUM$-ratio
should be also function of {\it density distribution} over the body.
If we confine ourselves by spherically-symmetric distribution of density $\rho(r)$, then
we have general expressions for the gravitational potential at radius $r$:
\be U(r)={G\,m(r)\over r}+\int_r^R\,4\,\pi\,G\,\rho(r)\,r\,dr;\ee
central potential:
\be U(0)=\int_0^R\,4\,\pi\,G\,\rho(r)\,r\,dr;\ee
potential energy: \be W={1\over 2}\,\int_0^M\,U(r)\,d m(r);\ee
and mass: \be dm=4\,\pi\,\rho(r)\,r^2\,dr;\quad
M=\int_0^R\,4\,\pi\,\rho(r)\,r^2dr.\ee
From these expressions we write down $WUM$ for spherical-summetrical heterogeneous bodies
as:
\be WUM={W\over U(0)\,M}={1\over 2} {\overline{U}\over U(0)};\quad 
\overline{U}={1\over M}\int_0^M\,U(m)\,dm.\ee
As a result, $WUM$-ratio of spherically-symmetric bodies is reduced to the ratio of the
mean value of monotonic function, $U(m)$,  to its particular value, $U(0)$, (ratio being
additionally divided by 2). The boundary values of this ratio can be found
pure mathematically for any given class of functions $\rho\,(r)$. We will not deal with
this abstract (though interesting) problem; instead, in the next  sections we consider two
cases of more or less {\it realistic} 
bodies, namely polytropes and two-phase spheres.
\section{Polytropes} \label{poly}
One case of {\it heterogeneous} bodies which apparently  should be considered first is the
case of classical polytropic stars. Using formulas from Chandrasekar's (1957)
classical text we have (some of these formulas are valid not only for
polytropic stars but we do not stop on these details) the next relations.\\
a){\it central potential} is expressed via other parameters of star as 
follows (Chan 100/85= Chandrasekhar (1957), p.100, Eq. (85)):
\be U_p(0)=(n+1){P_c\over \rho_c} + {G\,M\over R}.\ee
Here $n$  is {\it the polytropic index}, $P_c$ and $\rho_c$ are central
values of pressure and density, $M$ and $R$ are the mass and radius of the star.\\
The next formulas include the parameters of the Lane-Emden function (LEF) at the first
zero point: \be \ba{l} \xi=\xi_1,\quad \theta (\xi=\xi_1) \equiv \theta_1 = 0;\quad
\biggr[ {d\,\theta\,(\xi)\over d\,\xi} \biggr]_{\,\xi= \xi_1}
\equiv \,{\theta_1}'\,<\,0;\quad \mu_1=-\,\xi_1^2\,{\theta_1}'. \ea \ee
Central pressure is (Chan 99/80,81):
\be P_c={1\over 4\,\pi\,(n+1) ({\theta_1}')^2}{\,G\,M^2\over R^4} .\ee
Central density $\rho_c$ is related with the  mean density $\overline{\rho}$ of
the star as follows (Chan 78/99):
\be \rho_c={1\over 3}\quad {\xi\over -{\theta_1}'}\quad \overline{\rho}.\ee
Combining these formulas we get the final expression for the central
potential of polytropic star:
\be U_p(0)=\biggl( 1+ \quad {1\over
-\,\xi_1\,{\theta_1}'}\biggr){G\,M\over R}.\ee
b){\it potential energy}\\
We have for polyropic star the famous formula (Chan 101/90):
\be W_p={3\over 5-n}{G\,M^2\over R}.\ee c){\it WUM-ratio}: \be \label{WUMp}
WUM_p={3\over 5-n}\quad {1\over 1+ \, {\xi_1 \over \mu_1}}.\ee 
\begin{figure}  \label{POLfig} \includegraphics[scale=.6]{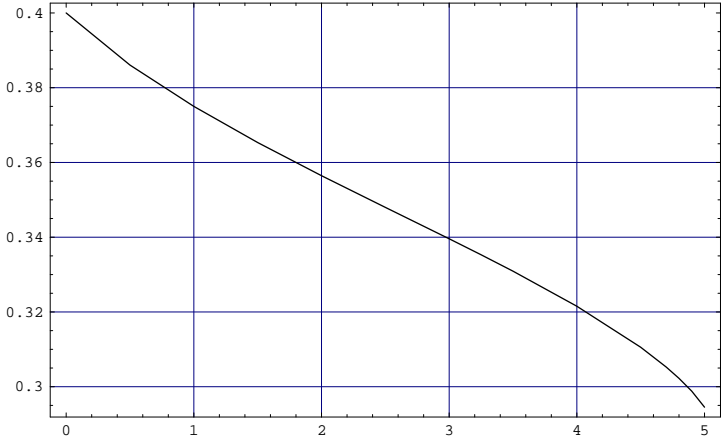}
 \caption{$WUM$ for polytropes.  Abscissas are values of polytropic index
$n$ and ordinates are values of the $WUM$-ratio.}  \end{figure} 
\subsection{$WUM$-ratio for polytropes}
Using values of parameters of polytropic stars given in 
(Chan, Table 4, p. 96) we calculated the values of $WUM$-ratio for 
values of polytropic index $n$ from $0$ to $4.9$, see Fig. 3.
However the limit  $n\rightarrow 5$ and in general region of values 
of $n$ close to 5 should be considered separately. First we note that
at $n\rightarrow 5$, $\quad \xi_1\rightarrow \infty$ and
there is indeterminacy of $\infty\over\infty$ kind in formula for 
$WUM_p$ Eq. (\ref{WUMp}). To solve this problem we use results of
Seidov and Kuzakhmedov (1979) who obtained, in particular, the dependence
 of $\xi_1$ for $n$ close to 5:
\be (0\,<\,(5-n)/5\,<<\,1),\quad \xi_1\,=\,{32\,\sqrt{3}\over \pi}\,\,
{1\over 5-n}.\ee  
Also it was shown by Seidov and Kuzakhmedov (1979)  that for values of $n$ close to 5, the
parameter $\mu_1$ is an {\it increasing} function of $n$:
\be \mu_1=\sqrt{3}\,(1+{1\over 12}\,\,{1\over 5-n}),\ee
that means that $\mu_1$ has minimum at $n$ about 4.82. This may  or may not lead to
minimum of $WUM_p$ as function of $n$. At point $n=5$ using LEF of index 5:
\be n=5,\quad \theta(\xi)=(1+{1\over 3}\,\xi^2)^{-1/2},\ee we get 
\be n=5,\quad WUM_p\,=\,{3\over 32}\,\pi =.294524.\ee 
Additionally we have two other analytical results for $WUM$
of polytropes : at $n=0$, $\,WUM=2/5$ and at $n=1$, $\,WUM=3/8$.
We recalculated the parameters of polytropes in the interval of
$n$ from 4 to 5 and found no minimum for $WUM$, see Fig. 3.
However we note that our boundary values differ from ones calculated by
Jabbar (1993) in the sense that ours are less than his. 
As one example, at $n=4.7$ Jabbar gives $\xi_1=54.810686$ as zero of LEF,  while our
calculations using Mathematica's command NDSolve give $\theta(54.810686)=-4.59 10^{-7}$
and our value of $\xi_1$ is in interval between $\xi=54.8098$ (at this point
$\theta=4.98*10^-8$) and $\xi=54.8099$ (at this point $\theta<0$). In general our values
of $\xi_1$
are less than Jabbar's. \section{Two-phase sphere} \label{2ph}
If $\rho_1$ and $\rho_2=q\,\rho_1$ are densities in envelope and core of sphere,
and $R$ and $r=x\,R$ are total radius of sphere and radius of core, then we have
next relations:\\ a){\it gravitational potential at the center}:
\be \label{U2ph} U_{2ph}(0)=2\,\pi\,G\,\rho_1\,R^2\,[1+(q-1)\,x^2];\ee
b){\it potential energy}:
\be \label{W2ph} W_{2ph}={16\,{\pi}^2\,G \over 15}{\rho_1}^2\,R^5\,
[1+{5\over2}(q-1)\,x^3+(q-1)(q-{3\over 2})\,x^5];\ee c){\it total mass}:
\be \label{M2ph} M_{2ph}={4\,\pi\over 3}\,\rho_1\,R^3\,[1+(q-1)\,x^3]; \ee
d)$WUM\,-\,ratio$: \be \label{WUM2ph} \ba{l} WUM_{2ph}={2\over 5}
{1+{5\over2}(q-1)\,x^3+(q-1)(q-{3\over 2})\,x^5 \over [1+(q-1)\,x^2]\,[1+(q-1)\,x^3]}.\ea \ee
\begin{figure}  \label{2PHfig} \includegraphics[scale=.6]{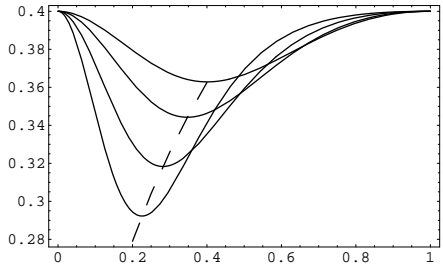}
 \caption{$WUM$ for two-phase spheres.  Abscissas are values of relative
radius of core, and ordinates are values of the $WUM$-ratio. Dash line
shows the locus of minimuma of $WUM$. } \end{figure} 
In Fig.4 we present dependence of $WUM$-ratio, for two-phase spheres with
various values of $q$, as function of {\it relative radius of core} $x=r/R$. Curves from
uppper one to lower one correspond to values of $q=3,\,5,\,10,\,$ and $20$, respectively. 
Both values of minimuma of $WUM$ and their "places", corresponding values of $x$, are
decreasing functions of $q$.
For values of $x$ corresponding to minimuma of $WUM$ the next simple equation is valid:
\be 6\,q^2\,x^5+5\,q\,(1-x)\,x^3\,(1+2\,x)-(1-x)^3\,(4+7\,x+4\,x^2)=0.\ee  or
\be q=\frac{(1 - x ) \,\left( \sqrt{96/x + 97 + 28\,x + 4\,x^2} -5 - 
10\,\sqrt{x}\right) } {12\,x^2}.\ee We used these formulas to calculate the dash line
 in Fig. 4.
\section{Stepenars} \label{step}
Here we briefly consider the case of simple spherically-symmetric density
distribution law allowing analytical  expression for WUM. We take:
\be \label{rost} \rho(r)=\rho_c\,(1-r/R)^{\nu},\ee
 where $\rho_c$ is central density and $\nu$ is a free parameter. By historical reasons we
refer to the gravitating bodies with density distribution  (\ref{rost}) 
as "stepenars", (Seidov, Kasumov, and Guseinov 1971).\\ We have:\\ {\it mass}:
\be M={8\, \pi\, \rho_c R^3\over (1+\nu)(2+\nu)(3+\nu)};\ee
{\it mean-to-central density ratio}: \be {\overline{\rho}\over \rho_c}={6\over
(1+\nu)(2+\nu)(3+\nu)};\ee \\{\it central potential}: 
\be U(0) =  {4\, \pi\,G\, \rho_c R^2\over (1+\nu)(2+\nu)};\ee
{\it potential energy}: \be W =
 {8\, \pi^2\,G\, \rho_c R^2\,(8+5\,\nu)\over
(1+\nu)^2(2\,+\nu)^2(3+2\,\nu)(5+2\,\nu)};\ee {\it WUM-ratio}:
\be WUM_{\nu} =  {(3+\nu)(8+5\,\nu)\over 4\,(3+2\,\nu)(5+2\,\nu)}.\ee
\begin{figure}  \label{STEPfig} \includegraphics[scale=.6]{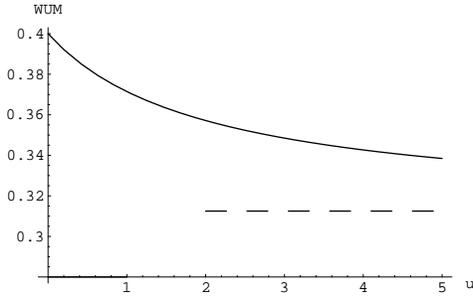}
 \caption{$WUM$ for stepenars.  Abscissas are values of  parameter $\nu$ in density
distribution law  $\rho(r)=\rho_c\,(1-r/R)^{\nu}$, and ordinates are values of
$WUM$-ratio. Broken line shows asimptotic value of $WUM\rightarrow 5/16$ at 
$\nu\rightarrow \infty$. } \end{figure} 
In Fig. 5 the dependence of $WUM$-ratio for stepenars as  function of
parameter $\nu$ is shown. In spite of large variation of matter concentration from 
$\rho=const$ at $\nu=0$, to $\overline{\rho}/ \rho_c\rightarrow 0$ at
$\nu\rightarrow\infty$,  $WUM$-ratio again as in polytrope's case lies in rather limited
interval from $2/5$ to $5/16$. \section{Alphars} \label{alpha}
Here we consider another example of "exotic" but simple density distribution, Seidov and 
Seidova (1971):
\be \label{roal} \rho(r)=\rho_c\,(r/R)^{\alpha},\quad 0\leq\alpha\leq 2. \ee We have:
\be WUM_{\alpha}={2-\alpha\over 5-2\,\alpha}.\ee \section{Discussion}
There is a rather classic problem of looking for general theorems of stellar 
structure, see e.g. chapter 3 in the classic text  Chandrasekhar (1957). 
The problem considered in this paper may be also referred to as  that dealing with 
general structure of celestial self-gravitating bodies.\\  We start from
interesting observation on one constant ratio, namely, (potential energy W)/((central
potential U ) x (total mass M)), in homogeneous ellipsoids and then try to look for
behavior of this ratio for another homogeneous bodies: rectangular parallelelepipeds and
symmetrical lenses. We found that in both cases $WUM$-ratios are confined in rather narrow
interval. Suprisingly, dependence of $WUM$  for  homogeneous rectangular parallelepipeds
(RP) on edge lengths ratio is non-monotonic: it has minimal value $.395437$ for cube while
any deviation from cube form to prolate RP (one dimension being smaller than two others)
or  elongated RP (one dimension being larger that two others) leads to the increase of 
value of $WUM$. In this respect the behavior of homogeneous rectangular
parallelepipeds is quite unlike the behavior of homogeneous ellipsoids
and there  is still some  mystery even to authors.\\
As to the homogeneous symmetrical lenses (SL) by Kondrat'ev and  Antonov (1993), here the
dependence of $WUM$ on parameters of SL is monotonic, however in this case there is
also some suprise  in the sense that  in the limiting case of thin symmetrical $spherical$
lens the $WUM$-ratio's value, ($128/105\,\pi$),  differs radically from the case of the
infinitesimally thin $quadratic$ plate with $WUM=1/2$.\\  Then we look for the
non-homogeneous however  spherically symmetric bodies and found that for the polytropes
with polytropic index $n$ in the  interval $0-5$, $WUM$-ratio again lies in narrow
interval from $2/5$ to $3/32\,\pi$. However for two-phase sphere with large ratio of
densities
$q=\rho_2/\rho_1$ it is possible to get very small  values of WUM. The physical reason of
it is that if we put in the center of any spherical symmetric star a (very) small but
dense spherical body then central potential may be very large while total potential energy
of star, being integral value, increases not so drastically. The effect of the strong
variation of density in the center of star, the "first-order phase-transition", is known
since pioneer works of W.H. Ramsey (1950).\\  In last two paragraphs of paper we consider
pure mathematical toy models in further attempts to  understand the behavior of the
$WUM$-ratio. We conclude  this discussion with notice that central-to-surface potential
ratio $U(0)/U(R)$ (among other "global" characteristics of the celestial self-gravitating
configurations) is also worth studying. For polytropes, $U(0)/U(R)=1+\xi_1/\mu_1$, see
section \ref{poly}.   \section*{Acknowledgements} We are
grateful to  Dr. E. Liverts for valuable discussions.
 \end{document}